# Simulation of the atomic structure near voids and estimation of their growth rate anisotropy

Andrei V. Nazarov[1,2], Alexander A. Mikheev[3], Aleksey P. Melnikov[1]

[1]National Research Nuclear University MEPhI, (Moscow Engineering Physics Institute), 31, Kashirskoe shosse 115409, Moscow, Russia.

[2] Institute for Theoretical and Experimental Physics named by A.I. Alikhanov of NRC "Kurchatov Institute", 25 Bolshaya Cheremushkinskaya str.,117218, Moscow, Russia

[3]The Kosygin State University of Russia, 33 Sadovnicheskaya str., 117997, Moscow, Russia.

We model the elastic field near the void using a new Molecular Static method. Results show that the atom displacements near nanovoids significantly differ for various crystallographic directions in bcc metals. Using some results of these calculations, we estimate the void growth rates following various directions basing on recently obtained equations of vacancy diffusion under strain. This growth rate anisotropy helps to explain the change in shape observed for many voids.

## 1. Introduction

Defects produced in materials by radiation damage are responsible for significant changes of their properties. To predict the lifetime of nuclear power plant components, one should understand the mechanisms of these changes. In particular, the formation and growth of voids and bubbles are of high interest for material performance in radiation environments [1-5]. Fundamental understanding of the void growth mechanism and kinetics is important for predicting the swelling kinetics of irradiated materials and changes of the materials strength properties [1-6]. In this work we focus on the void shape. Voids usually exhibit a well-defined, non-spherical shape [3-5,7]. Sometimes the cavities were first in the form of spheres, and later are transformed into to cubes bounded by {100}-facets [7]. Experimentally, voids in metals and alloys both with bcc [5,7] and fcc [7-9] and hcp [10] structure have non-spherical shape. In the numerous experiments during past three decades, the evolution of void morphologies during irradiation or post-irradiation annealing in various materials was studied [3-5,7-13]. However, a systematic study on the void shapes considering the different factors is still missing [7]. There are several influencing factors for the void shape development and its change during irradiation- in particular, surface energy anisotropy, growth anisotropy of void surfaces, a preferential adsorption of atoms at certain void surfaces [3], and diffusion anisotropy in metals [10]. In this paper we study the influence of another factor on the void growth anisotropy: the effect of an elastic field in the vicinity of voids on diffusion fluxes of vacancies and a shift of the void surface elements for various crystallographic directions.

Generally, displacement fields near spherical voids are determined by the equations of isotropic elasticity theory [14]. Solutions of these equations [14] show that the elastic field in the vicinity of voids does not contribute to the diffusion of vacancies [1] and does not affect the growth rate of the voids. Therefore, the voids are called neutral sinks for defects [1].

In the work, we first use the recently developed model [14] to study the structure in the vicinity of spherical nanovoids in bcc iron. After this, we study the vacancy diffusion near spherical voids and shifting rate anisotropy of the nanovoid surface element. Moreover, we believe that the anisotropy of the vacancy flux and void growth are conditioned by the anisotropy of the displacement field in the vicinity of the voids and discreteness of atomic structure in metals.

The paper consists of several sections: a model for simulation of an atomic structure in the region near nanovoids is described and summarized in Section 2; results are presented in Section 3; equations for the shift rate of the void surface elements for various crystallographic directions are derived in Section 4; the shifting rate calculations of the void surface elements are presented and discussed in more detail in Section 5 before conclusions are drawn in Section 6.

## 2. Molecular Statics model. New approach

In this section we describe the main features of the model that distinguish it from those generally used in the Molecular Statics method. The equilibrium positions of atoms in the main computational cell are simulated using a variational procedure, which is usually employed in the molecular statics method [18]. At the same time, we take into account interactions with atoms embedded in an elastic medium surrounding the main computational cell. In contrast to the previously developed model [19], we apply iterative algorithm of the self-consistent atomic coordinates calculation in the main cell and the calculated displacement field characteristics in the elastic medium [20-23].

The computational cell (zone I, Fig. 1) is a sphere ($Rs$) containing more than 300000 atoms. It is surrounded by atoms embedded into an elastic medium (zone II, Fig. 1). The displacements **u** of these atoms, generated by a void, are based on the first term of the static isotropic elastic equation solution [14,18,21]:

$$\boldsymbol{u} = \frac{C_1 \boldsymbol{r}}{r^3}, \tag{1}$$

The constant $C_1$ is calculated according to Eq. (1) by using the results of atomic displacements simulation in the computational cell for the atoms that are located in a spherical layer. The constant $C_1$ calculated at the previous step of the iterative procedure is used to determine atomic displacements of the elastic medium II. Then, the relaxation of the atoms of zone I is carried out anew, and the constant $C_1$ is calculated again. Our model in all have been described in detail [22,23].

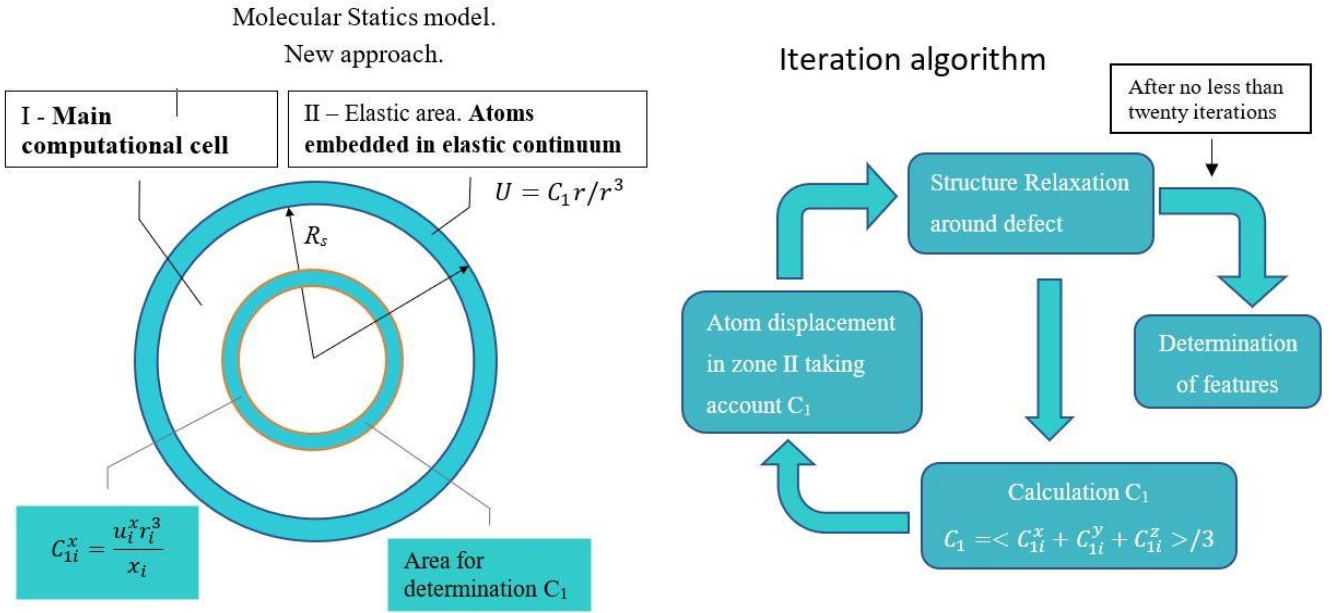

**Fig. 1.** An illustration of the simulation cell and the main steps of the iterative algorithm.

The model allows one to obtain not only the energy of formation and migration of defects, but also the characteristics that determine the effect of pressure on the concentration of point defects and their migration rate [20-23]. Our model also simulates the atomic structure in the vicinity of vacancy complexes [24], clusters [25 and voids [26]. The next section presents the simulation results for voids of two sizes in the bcc iron that are used further in the shifting rate calculations of the void surface elements.

## 3. Results of MS simulation

Simulations are performed using a well-examined embedded-atom method interatomic potential for bcc iron [27]. The simulation results for displacements of atoms in the vicinity of voids of the radius $R$ = 9.4 Å and size $R$ =14.3 Å in the directions <100>, <110>, <111> are presented on Fig. 2. and Fig. 3.

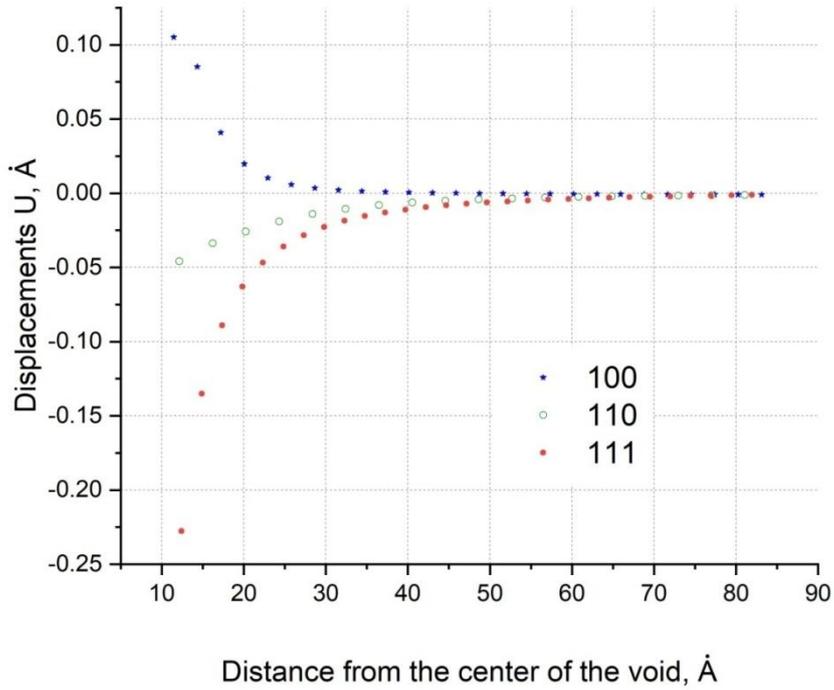

**Fig. 2.** The atom displacements for various crystallographic directions in bcc iron for nanovoid (R = 9.4 Å)

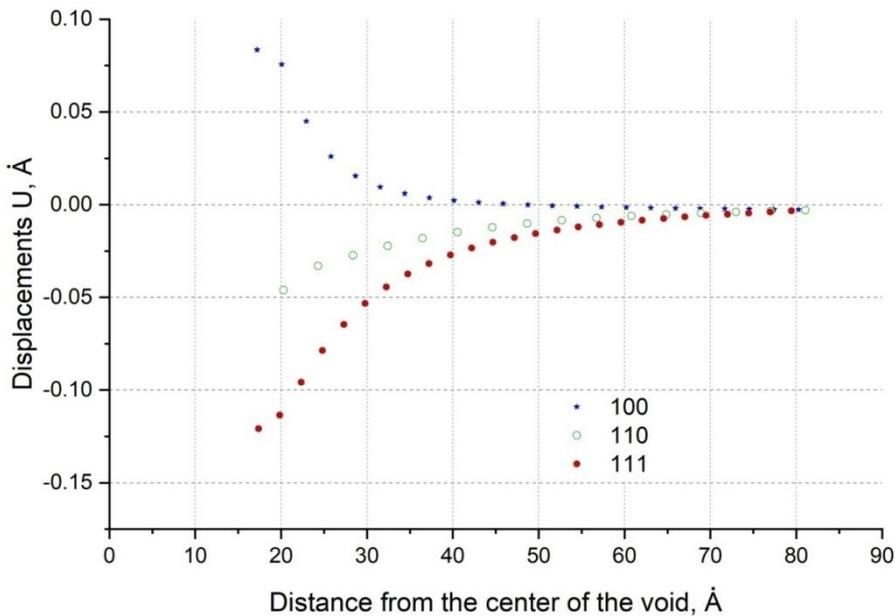

**Fig. 3.** The atom displacements for various crystallographic directions in bcc iron for nanovoid ($R$ =14.3 Å)

The data shown in the figures, as well as previously obtained for voids of other sizes [26], show that the atom displacements near nanovoids differ sufficiently from the elasticity theory solution and in addition, the displacements significantly differ for various crystallographic directions. Moreover, the displacements in the crystallographic directions <100> are positive at rather large distances from the surface of the voids.

Apparently, for the first time, a qualitatively analogous character of atomic displacements was discovered by modeling a structure in the vicinity of a vacancy in BCC iron in [18]. Similar results for vacancies in iron and other BCC metals are found in publications [20-23]. Also, in [25] we find similar structural features for symmetric clusters of vacancies. In the case of vacancies, these structural features significantly affect the estimations of the formation energy and migration energy, as well as the values of the formation volumes and migration volumes [22,23]. We assume that the mentioned features of the displacements are a consequence of the system discreteness with a bcc structure.

We further study how this feature that is become apparent in the simulation of spherical voids, affects the shifting rate of the void surface elements. Vacancy flux on the void surface determines growth rate of the voids [1,15,17]. In the traditional approach, it is assumed [1,17] that the contribution to the diffusion vacancy flux of the elastic field can be found using the solution of equations from the classical theory of elasticity in the vicinity of the void. Since in this case [15,17]

$$\nabla Sp\varepsilon = 0, \qquad (2)$$

there is no stress influence on flux of vacancies [15,17].

Obtained results show that

$$\nabla Sp\varepsilon \neq 0, \qquad (3)$$

and a kinetics "equation for the growth rate of voids must contain the additional terms conditioned by strains, arising from voids"[15].

## 4. Void growth rates for different directions

As it was done in Lifshitz-Slezov [LS] theory [17], the growth of the voids is considered by using the mean-field approximation and steady-state solution of the diffusion equation around each void. The growth rate of the void in the general case is determined by the equation [17]:

$$\frac{dV}{dt} = -\iint (\boldsymbol{n}, \boldsymbol{j}) ds, \qquad (4)$$

where $V$ is the volume of the void, $\boldsymbol{n} = \{n_x, n_y, n_z\}$ is the unit vector that is normal to the void surface, $\boldsymbol{j} = \{j_x, j_y, j_z\}$ is the diffusion flux vector, $(\boldsymbol{n}, \boldsymbol{j})$ is the scalar product, and $ds$ is the surface element.

The left-hand side of Eq. (4) after differentiation has a view:

$$\frac{4\pi R^2 dR}{dt} = -\iint (\boldsymbol{n}, \boldsymbol{j}) ds. \qquad (5)$$

It can be shown by a way that is similar to one applied by us earlier [28], that we can use the next equation for the shifting rate of a void surface element in a certain direction:

$$\frac{dR}{dt} = -(\boldsymbol{n}, \boldsymbol{j}). \qquad (6)$$

The x component of the vacancy flux in the zero approximation of the field effect, according to the results of [16] has the form:

$$j_x = -D_V \left[ \frac{\partial c}{\partial x} - c \frac{K^V}{kT} \frac{\partial Sp\varepsilon}{\partial x} \right], \qquad (7)$$

where $c$ is the vacancy concentration, $D_V$ is the vacancy diffusion coefficient in the perfect (unstrained) system and

$$\varepsilon_{ij} = \frac{1}{2} \left( \frac{\partial u_i}{\partial x_j} + \frac{\partial u_j}{\partial x_i} \right) \qquad (8)$$

is the strain tensor ($i, j = 1,2,3$), $Sp\varepsilon$ is the trace of the tensor,

$$K^V = \frac{1}{2} \sum_s \sum_{k \neq s} \frac{(x_{ks}^V)^2}{R_{ks}^V} \frac{\partial E}{\partial R_{ks}} \bigg|_{R_{ks}^V}, \qquad (9)$$

$x_k$, $y_k$, $z_k$ are the coordinates of the atom $k$ that can belong to zone I only (see Fig.1), $x_{ks} = x_k - x_s$, $y_{ks} = y_k - y_s$, $z_{ks} = z_k - z_s$, $k \neq s$, $x_s$, $y_s$, $z_s$ are the coordinates of the atom $s$ that can belong to zone I and zone II, under atom structure simulations near the vacancy [16]. $R_{ks} = |\mathbf{r}_k - \mathbf{r}_s| = \sqrt{x_{ks}^2 + y_{ks}^2 + z_{ks}^2}$ for all atoms $k$. The distances $R_{ks}^V$ are calculated using the coordinates of the system configuration including vacancy. It should be emphasized that the energy of the system $E$ (see Eq. (9)) depends on the distances $R_{ks}$ and does not depend on external elastic fields.

In order to evaluate the effect of the elastic field on the vacancy flux in region near void and obtain analytical solutions, we use the method of successive approximations. And as the zero[th] approximation, we choose the solution of the diffusion equation that does not take into account the effect of the field, as it was done in [17]. The concentration of vacancies inside a spherical layer having an internal radius $R$ and outer radius $R_G$ in the steady-state approximation has the form:

$$c = -\frac{RR_G(c_m - c_R)}{R_G - R}\frac{1}{r} + \frac{R_G c_m - R c_R}{R_G - R}, \qquad (10)$$

where $r$ is the distance from the center of the void $c_R = c(R)$, $c_m = c(R_G)$ and

$$c_R = c_{eq} \exp\left(\frac{2\gamma V^f}{kTR}\right), \qquad (11)$$

where $c_{eq}$ is the equilibrium concentration of the vacancies for a flat interface, $\gamma$ is the surface energy and $V^f$ is the formation volume of the vacancy.

The growth rate of the void in [17] is isotropic and has a view:

$$\frac{dR}{dt} = c_{eq} D_V \left[\frac{R_G}{R_G - R}\frac{1}{R}\left(\Delta + 1 - \exp\left(\frac{2\gamma V^f}{kTR}\right)\right)\right], \qquad (12)$$

where $\Delta = \frac{c_m - c_{eq}}{c_{eq}}$ is the supersaturation of the vacancies.

Now we evaluate shifting rate of the nanovoid surface element in the different directions taking into account the elastic field near the void.

*4.1. The shifting rate of void surface in the direction <100>*

There is one component of the diffusion flux vector for this crystallographic direction in the coordinate system, associated with the crystallographic axes:

$$j_x = -D_V\left[\frac{\partial c}{\partial x} - c\frac{K^V}{kT}\frac{\partial Sp\varepsilon}{\partial x}\right] \qquad (13)$$

and the scalar product in Eq. (6) has only one terms:

$$(\mathbf{n},\mathbf{j}) = j_x. \qquad (14)$$

The derivatives of the concentrations (see Eq.(10)) at each point of the spherical layer depend on the radius $r$ and coordinates in this point:

$$\frac{\partial c}{\partial x} = \frac{RR_G(c_m - c_R)}{R_G - R}\frac{x}{r^3}, \quad \frac{\partial c}{\partial y} = \frac{RR_G(c_m - c_R)}{R_G - R}\frac{y}{r^3}, \quad \frac{\partial c}{\partial z} = \frac{RR_G(c_m - c_R)}{R_G - R}\frac{z}{r}. \qquad (15)$$

As a result, the concentration derivative takes the form:

$$\frac{\partial c}{\partial x} = \frac{c_{eq} RR_G}{R_G - R}\frac{x}{r^3}\left[\Delta + 1 - \exp\left(\frac{2\gamma V^f}{kTR}\right)\right]. \qquad (16)$$

The concentration derivative at the pore surface at $x = R$, $r = R$ is described by the equation:

$$\frac{\partial c}{\partial x} = \frac{c_{eq} R_G}{R_G - R} \frac{1}{R} \left[ \Delta + 1 - exp\left(\frac{2\gamma V^f}{kTR}\right) \right]. \tag{17}$$

After mathematical transformations, we obtain a kinetic equation for shifting rate of the void surface element in directions <100>:

$$\frac{dR}{dt} = c_{eq} D_V \left[ \frac{R_G}{R_G - R} \frac{1}{R} \left( \Delta + 1 - exp\left(\frac{2\gamma V^f}{kTR}\right) \right) - \frac{K^V}{kT} exp\left(\frac{2\gamma V^f}{kTR}\right) \frac{\partial Sp\varepsilon}{\partial x} \right]. \tag{18}$$

*4.2. The shifting rate of void surface in the direction <110>*

It is necessary to calculate the scalar product of the flux vector by a unit vector orthogonal to the void surface to obtain an equation for the shifting rate of void surface in the direction <110>. As the flux vector is given by:

$$\boldsymbol{j} = \{j_x, j_y, 0\}, \tag{19}$$

the unit normal vector to the void surface has a view:

$$\boldsymbol{n} = \left\{\frac{1}{\sqrt{2}}, \frac{1}{\sqrt{2}}, 0\right\}, \tag{20}$$

and the scalar product in Eq. (6) has two terms:

$$(\boldsymbol{n}, \boldsymbol{j}) = \frac{j_x}{\sqrt{2}} + \frac{j_y}{\sqrt{2}}. \tag{21}$$

At the void surface $x = \frac{R}{\sqrt{2}}$, $y = \frac{R}{\sqrt{2}}$, $r = R$, and if we take into account Eq. (15) we obtain:

$$\frac{\partial c}{\partial x} + \frac{\partial c}{\partial y} = \frac{R_G (c_m - c_R)}{R_G - R} \frac{\sqrt{2}}{R}. \tag{22}$$

If we use Eqs. (11) and (22), then after mathematical transformations the kinetic equation for shifting rate of the void surface element in the directions <110> takes the form:

$$\frac{dR}{dt} = \frac{c_{eq} D_V}{\sqrt{2}} \left[ \frac{R_G}{R_G - R} \frac{\sqrt{2}}{R} \left( \Delta + 1 - exp\left(\frac{2\gamma V^f}{kTR}\right) \right) - \frac{K^V}{kT} exp\left(\frac{2\gamma V^f}{kTR}\right) \left(\frac{\partial Sp\varepsilon}{\partial x} + \frac{\partial Sp\varepsilon}{\partial y}\right) \right]. \tag{23}$$

*4.3. The shifting rate of void surface in the direction <111>*

Now it is necessary to calculate the scalar product of the flux vector by a unit vector orthogonal to the void surface to calculate the shifting rate of void surface in the direction <111>. As the flux vector is given by:

$$\boldsymbol{j} = \{j_x, j_y, j_z\}, \tag{24}$$

the unit normal vector to the void surface has a view:

$$\boldsymbol{n} = \left\{\frac{1}{\sqrt{3}}, \frac{1}{\sqrt{3}}, \frac{1}{\sqrt{3}}\right\}, \tag{25}$$

and the scalar product in Eq. (6) has three terms:

$$(\boldsymbol{n}, \boldsymbol{j}) = \frac{j_x}{\sqrt{3}} + \frac{j_y}{\sqrt{3}} + \frac{j_z}{\sqrt{3}}. \tag{26}$$

At the void surface $x = \frac{R}{\sqrt{3}}$, $y = \frac{R}{\sqrt{3}}$, $z = \frac{R}{\sqrt{3}}$, $r = R$, and if we take into account Eq. (15) we obtain:

$$\frac{\partial c}{\partial x} + \frac{\partial c}{\partial y} + \frac{\partial c}{\partial z} = \frac{R_G(c_m - c_R)}{R_G - R} \frac{\sqrt{3}}{R}. \tag{27}$$

If we use Eqs. (11) and (27), then the kinetic equation for shifting rate of the void surface element in the directions <111> takes the form:

$$\frac{dR}{dt} = \frac{c_{eq} D_V}{\sqrt{3}} \left[ \frac{R_G}{R_G - R} \frac{\sqrt{3}}{R} \left( \Delta + 1 - exp\left(\frac{2\gamma V^f}{kTR}\right) \right) - \frac{K^V}{kT} exp\left(\frac{2\gamma V^f}{kTR}\right) \left( \frac{\partial Sp\varepsilon}{\partial x} + \frac{\partial Sp\varepsilon}{\partial y} + \frac{\partial Sp\varepsilon}{\partial z} \right) \right]. \tag{28}$$

Note that the Eqs. (18), (23), (28) allow us to estimate the effect of the surface energy anisotropy on the shift rate of the void surface elements. Moreover, it is easy to see that similar exponential dependences on this parameter are present in the first and second terms in square brackets. However, for now we will not take into account this effect, since it is more rational to do this in the framework of a more complete model, the main features of which are outlined at the end of the next section.

## 5. Results and discussion

Then after the polynomial approximation of the components of the atom displacement vectors for the selected crystallographic directions obtained in the simulation we calculate the corresponding components of the strain tensor and coordinate derivatives of the strain tensor trace. It allows to calculate the dependences of the shifting rate of the void surface elements for selected crystallographic directions from temperature for different values of supersaturations with help the Eqs. (18), (23) and Eq. (28). We used the values of $V^f$ and $K^V$ that were previously calculated in our work [23].

Fig. 4 and Fig. 6 show the above-mentioned dependences normalized to the product of the vacancy equilibrium concentration and the vacancy diffusion coefficient at the corresponding temperature. In addition Figs. 5 and 7 show the ratios of the calculated dependences to analogous ones, but without taking into account the influence of the elastic field in the vicinity of the voids [1, 17].

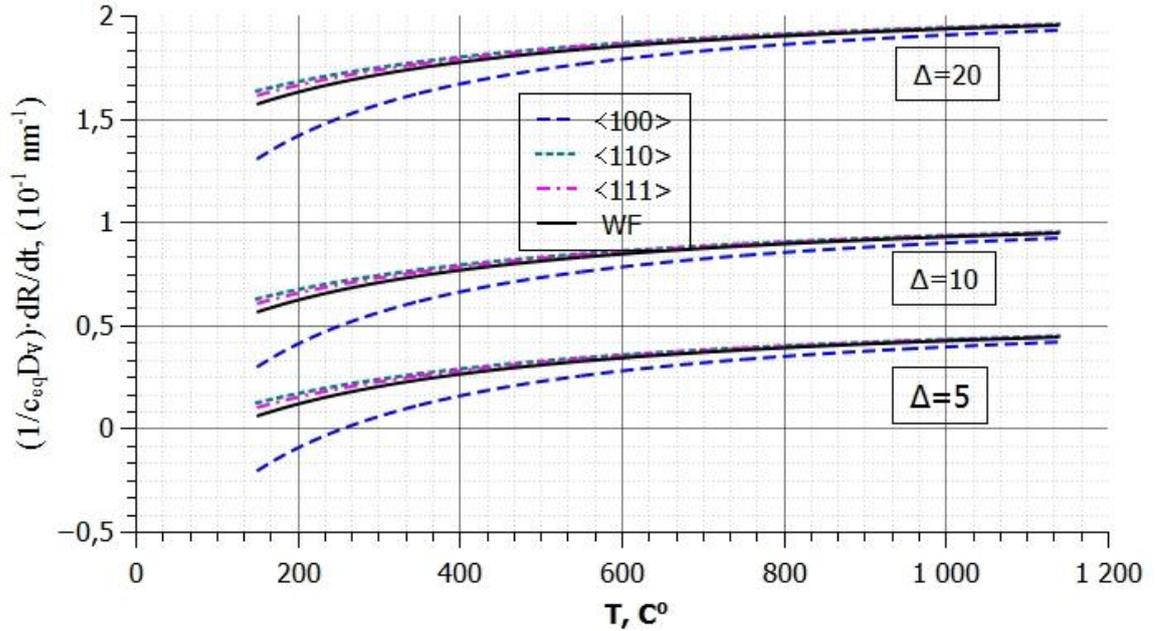

**Fig. 4.** The void growth rates for different directions at some values of supersaturation. $R = 0.94$ nm

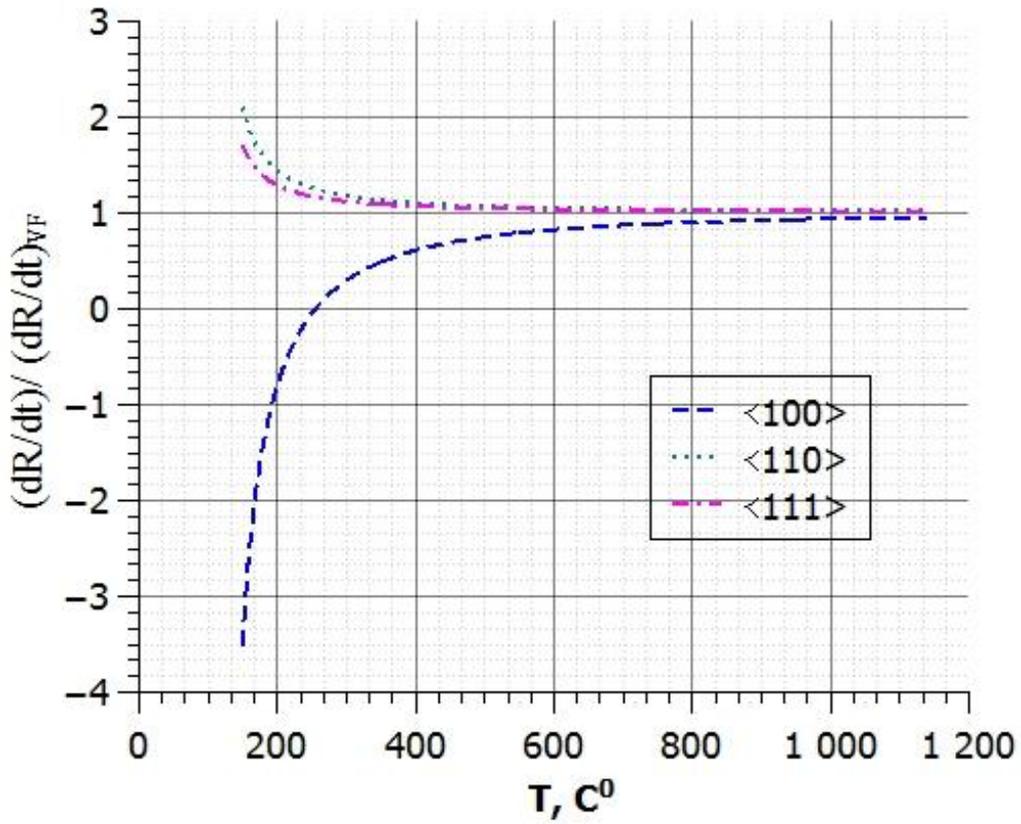

**Fig. 5.** Normalized void growth rates for different directions. $\Delta = 5$, $R = 0.94$ nm

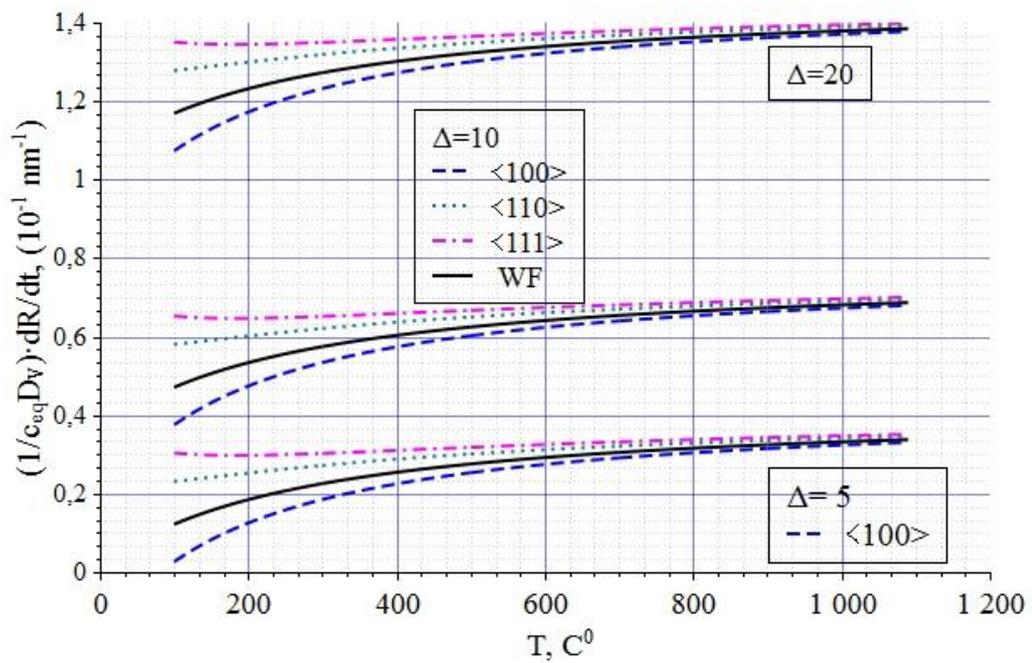

**Fig. 6.** The void growth rates for different directions at some values of supersaturation. $R = 1.43$ nm

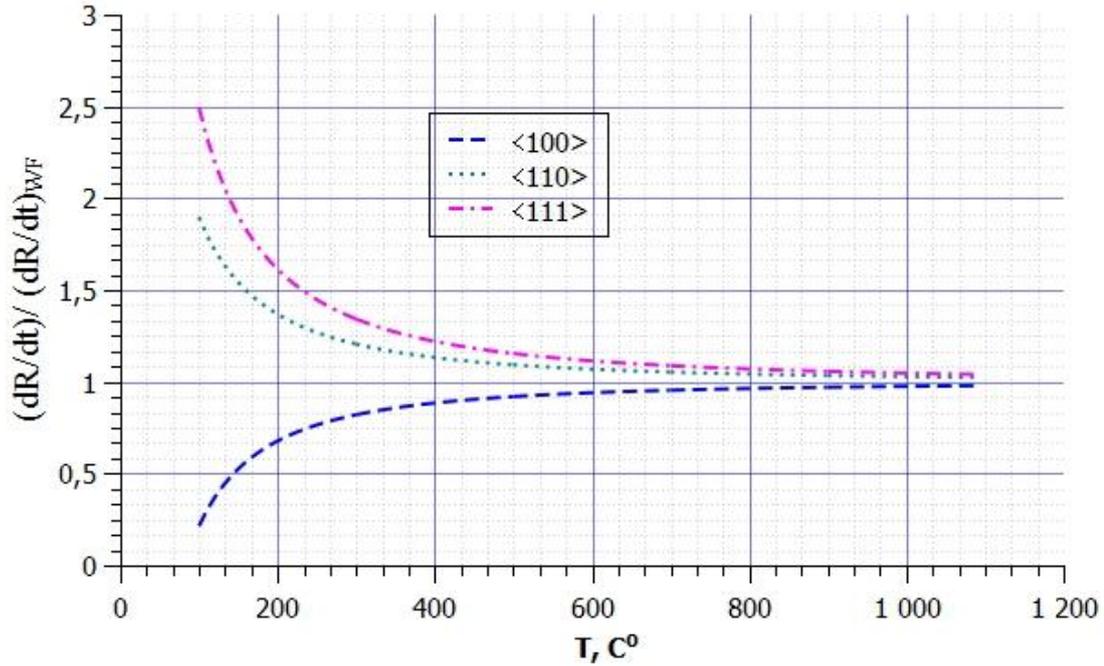

**Fig. 7.** Normalized void growth rates for different directions. $\Delta = 5$, $R=1.43$ nm

These data show, that in the case of nano-sized voids the displacement asymmetry of atoms in the vicinity of voids in bcc iron leads to a significant slowing shifting of the surface elements in directions that coincide with the crystallographic <100> type, or close to them.

Thus, the effect found in this work is very likely to be one of the reasons for the appearance of voids having a non-spherical but cuboid form. These voids were repeatedly observed in the experiments in irradiated materials with a bcc structure [5,12].

Note that in work [15] we used generalized diffusion equations for vacancy flux [16] that takes into account the influence of the elastic field on atomic configurations of the defect environments when the atom arrives to the saddle point. These equations allow to describe the influence of elastic strain near the voids more accurately. As result a kinetic equation for the growth rate of voids contains additional terms due to strains arising near voids. This feature distinguishes the kinetic equation from one that is known in Lifshitz and Slezov (LS) theory [17] and changes the kinetic of void growth. Thus, elastic strain in the vicinity of the nanovoids affect both the void growth rate and their shape change.

Recall that we proceeded on the assumption that at the initial instant of time the void has a spherical shape. It is clear that the complete model of void growth must take into account consistently all stages and proceed from an agreed system of equations or models:

1. At the first we have to solve the diffusion equation numerically, taking into account the influence of the elastic field.
2. Then it is necessary to calculate the rates of the shifting of the void surface elements.
3. At this stage it is necessary to calculate the change in the shape of the void in a short interval of time $\Delta t$ on the basis of the rates of surface element shifting.
4. Finally, we must simulate the atomic structure in vicinity of the void knowing its shape.

After that we must go to step 1.

By executing this sequence, one can obtain a picture of the change in the shape of the void with time for given external parameters (primarily supersaturation and temperature).

## 6. Conclusions

- New model for determining atomic structure in the vicinity of nanovoids is presented.
- The obtained fields of displacements for nanovoids are sufficiently different from the elasticity theory solution.

- Also, the displacements are significantly different for various crystallographic directions in metals with bcc structure.
- The shift rate equations of the void surface elements are obtained for various crystallographic directions
- The shift rate of the void surface elements is calculated for various crystallographic directions in a wide range of temperatures.
- Consideration of the asymmetric elastic field effect in the vicinity of the pore leads to the conclusion that the strain field significantly slows down the shift rate of the pore surface elements in the direction coinciding with and close to <100> crystallographic one.
- This effect should lead to a change in the shape of the initially spherical void and, probably, gradually makes the shape closer to the cuboid type.

**Acknowledgements**

Authors would like to acknowledge the financial support of the National Research Nuclear University MEPhI Academic Excellence Project (Contract No. 02.a03.21.0005).